\documentclass[journal]{IEEEtran}

\ifCLASSINFOpdf
 \usepackage[pdftex]{graphicx}
 \usepackage[justification=justified]{caption}
 \graphicspath{{../pdf/}{../jpeg/}}
 \DeclareGraphicsExtensions{.pdf,.jpeg,.png}
\else
 \usepackage[dvips]{graphicx}
\graphicspath{{../eps/}}
\DeclareGraphicsExtensions{.eps}
\fi

\ifCLASSOPTIONcompsoc
\usepackage[caption=false,font=normalsize,labelfont=sf,textfont=sf]{subfig}\else
\usepackage[caption=false,font=footnotesize]{subfig}
\fi

\usepackage{fixltx2e}
\usepackage{romannum}
\usepackage{stfloats}

\begin{document}

\title{A Multi Polarization Square Patch Antenna with a Reconfigurable Feeding Network}

\author{Roozbeh~Rezaeipour,~\IEEEmembership{Student,~K.N. Toosi University of Technology}
       \\ Ramezanali~Sadeghzadeh,~\IEEEmembership{Professor of Electrical Engineering,~K.N. Toosi University of Technology}}


\maketitle

\begin{abstract}
A multi-polarization square patch antenna with a reconfigurable feeding network is presented in this paper. The reconfigurable feeding network of this antenna is implemented on a FR-4 substrate by a Wilkinson power divider and a branch line coupler which perform amplitude distribution in the feeding network. Besides, two switching circuits which consist of one PIN diode (BAR63-02w) and its DC biasing circuit manage the RF signal flow on this feeding network. These switching circuits control the phase of the RF signal applied to the square patch, so it can provide linear polarization, left-hand and right-hand circular polarization at 2.45 GHz which has many applications in wireless networks. The simulated and measured results are presented which illuminate acceptable axial ratio bandwidth (ARBW) for both right-hand and left-hand circular polarization in (2.38-2.48 GHz) and minimum -10 dB return loss at 2.45 GHz.

\end{abstract}


\IEEEpeerreviewmaketitle

\section{Introduction}

\IEEEPARstart{M}{Multi}-polarization antenna has appealed attentions since compact wireless systems were expanded. These antennas are also known as “polarization agile antenna” which is used in Wireless Local Area Network (WLAN), microwave tagging systems, and diversity polarization. Their characteristics can be altered in real-time and be capable of communication system considerations. Hence, they have a lightweight and compact size which is essential for wireless systems. Furthermore, reusing of the frequency spectrum and doubled capacity are satisfied in such designs \cite{r7}\cite{r11}. Multi-polarization can relieve signal losing led by multi path effect. It can be implemented in two ways: using reconfigurable radiation elements or designing suitable feeding networks \cite{r5}.

The reconfigurability is implemented in the main characteristics of an antenna by manipulating the current distribution over the antenna element or feeding network structure. A variety of methods are utilized for the reconfigurable antenna. \cite{r13} - [16]. In \cite{r13}, dual-frequency operation is satisfied by locating varactors and capacitors in slots. The second resonant frequency can be tuned by changing the DC voltage. Also in \cite{r12}, a hybrid reconfigurable antenna is introduced which frequency reconfigurability is realized by controlling the states of switch electronically.  Moreover, pattern-reconfigurable antenna is the other way of reusing common structures of an antenna which has been explored in \cite{r14}, \cite{r30}. In \cite{r14}, pattern-reconfigurable antenna is achieved over a square-ring patch antenna by placing four shorting walls and controlling two numbers of walls by applying DC voltage so antenna can operate in two modes: monopolar plat-patch and normal patch modes. Hence, antenna’s radiation pattern can be switched between conical and broadside radiations pattern at a fixed frequency. Additionally, polarization reconfigurable antennas based on two ways of implementation is presented in \cite{r22}-[16]. 

Using active microwave elements such as PIN diodes, varactors, RF switching circuits or power dividers are frequent in polarization reconfigurable antenna designing. Two methods are employed for this type of antenna which are using the active component as a part of radiation elements of antenna structure or using them in feeding network. In both methods, switches can be controlled in ON and OFF states by a DC controlling circuit \cite{r20}. In \cite{r22}, two slot-ring with perturbations antenna is presented which can switch between two CP and one LP by creating discontinuities at specific angle and placing PIN diode for switching polarization. Perturbation is used in \cite{r11} on a square-ring slot. Small conducting pads are connected to each other by PIN diode and multi-polarization is implemented trough diode switching. In \cite{r24}, two PIN diodes added in the optimized location of the E-shaped antenna structure. The antenna has simple structure that is capable of switching its polarization between RHCP and LHCP. In \cite{r25}, a simple microstrip patch antenna with a slot fed by a coaxial diagonally is probed. Pin diode is in charge of adjusting CP polarization in different frequency by two capacitors and DC controlling block. In \cite{r29}, \cite{r19}, same structure is introduced with one and two feed line respectively. In square patch antenna, circular polarization is realized by truncating corners of the patch \cite{r33}. By placing pin diodes with their DC bias on cut corners multi-polarization is achieved on this structure. 
Altering in feeding networks is performed in \cite{r2}-\cite{r17}. In \cite{r2}, multi-polarization is designed by a four-way power divider to feed the antenna. When all patches are excited with equal phases and amplitudes linear polarization is radiated whereas equal amplitudes and 90 degrees phase differences for each output conduct circular polarization. Although in \cite{r16} and \cite{r17} feeding networks are sophisticated, all structures led a multi-polarization antenna with 4 polarizations. PIN diodes are used in feeding network design and their DC controlling block can select suitable polarization for radiating at 2.4 GHz.All mentioned antennas utilize RF components in their structure for achieving reconfigurability.

In this paper, a reconfigurable feeding network is implemented by two types of power dividers and pairs of PIN diodes. Power dividers perform dividing power applied to the input port of the antenna equally between two ports connected to the microstrip patch antenna. Pairs of PIN diodes and their DC controlling circuits manage the phase of the RF signal going through a branch line coupler to the microstrip patch antenna. As a result, a multi-polarization antenna will be realized on this structure.
The desirable antenna structure with its feeding network and explanation of the components' function is introduced in section \Romannum{2}. Then, in section \Romannum{3} results of the simulation done by CST are presented. Moreover, experimental results are presented in \Romannum{4}. Finally, section \Romannum{5} is the conclusion.

\section{Antenna Configuration and Design Process}
\begin{figure}[!t]
\centering
\includegraphics[width=2.5in]{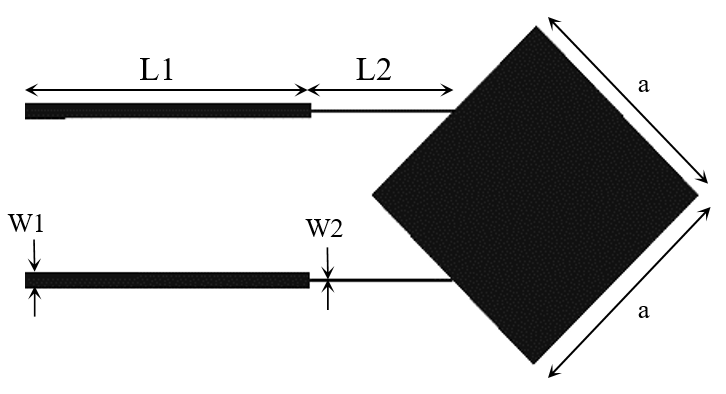}
\caption{Structure of the antenna}
\label{fig_sim}
\end{figure}

\begin{table}[!t]
\renewcommand{\arraystretch}{1.3}
\caption{Dimension of Proposed Antenna in mm}
\label{table_example}
\centering
\begin{tabular}{|c|c|c|c|c|}
\hline
a & W1 & L1 & W2 & L2 \\
\hline
27 & 1.84 & 36 & 0.26 & 18\\
\hline
\end{tabular}
\end{table}

\subsection{Antenna Configuration}
Technically, achieving circular polarization on a microstrip patch antenna is based on two factors: equal amplitude and   phase difference between excited modes of the antenna. Moreover, there are many methods for realizing circular polarization on microstrip patch antenna rooted in aforementioned factors [17]. In this paper, a square microstrip patch antenna (Fig. 1) is designed for 2.45 GHz and fabricated on a FR4 substrate with 1 mm height and 4.6 permittivity. Circular polarization is implemented by exciting central points of two edges of the square microstrip patch antenna by equal amplitude and 90 degrees phase difference. As a result, TM01 and TM10 modes of this structure are excited with 90 degrees phase difference leading to radiate circular polarization (CP). In addition, linear polarization (LP) is accomplished on the same design by feeding the square patch antenna with identical amplitude without phase difference. The parameters and dimensions of the antenna and matching lines are handed in Table \Romannum{1}.

\subsection{Feeding Network}
Power dividers are broadly used in feeding networks. In this paper, a Wilkinson power divider (Fig. 2) and a branch line coupler (Fig. 3) are utilized as feeding network of the proposed antenna. Parameters and dimensions of these power dividers are given in Table \Romannum{2} and \Romannum{3} respectively. Feeding network of this antenna is designed for 2.45 GHz and constructed on a FR-4 substrate with 4.6 permittivity. In first part of this feeding network, a RF signal applied to the Wilkinson power divider lead equal phases and amplitudes at output ports of this power divider. In the second part of this feeding network, the RF signal enters the branch line coupler which can turn applied power into two RF signals with equal amplitude and 90 degrees phase difference. Consequently, if aforementioned antenna in section A. is joined to this feeding network, circular polarization’s conditions will be satisfied. On the other hand, if both ports of the branch line coupler are excited, the antenna will be fed by a RF signal with equivalent phase and amplitude which is indispensable for linear polarization.

\begin{figure}[!t]
\centering
\includegraphics[width=2.5in]{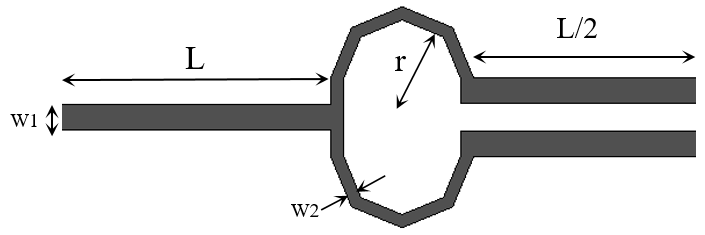}
\caption{Structure for the Wilkinson power divider}
\label{fig_sim}
\end{figure}

\begin{table}[!t]
\renewcommand{\arraystretch}{1.3}
\caption{Dimension for the Wilkinson power divider in mm}
\label{table_example}
\centering
\begin{tabular}{|c|c|c|c|c|}
\hline
r & L & W1 & W2 \\
\hline
4.72 & 36 & 1.84 & 0.96\\
\hline
\end{tabular}
\end{table}

\begin{figure}[!t]
\centering
\includegraphics[width=2.5in]{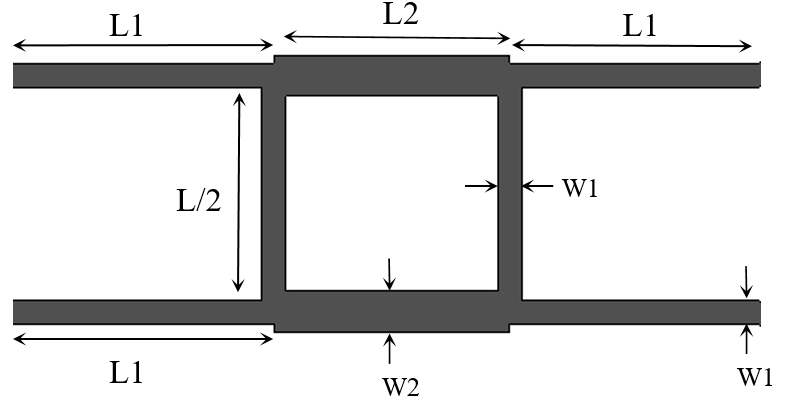}
\caption{Structure of the branch line coupler}
\label{fig_sim}
\end{figure}

\begin{table}[!t]
\renewcommand{\arraystretch}{1.3}
\caption{Dimension of the branch line coupler in mm}
\label{table_example}
\centering
\begin{tabular}{|c|c|c|c|c|}
\hline
L1 & W1 & L2 & W2 \\
\hline
36 & 1.84 & 16.16 & 0.96\\
\hline
\end{tabular}
\end{table}

\begin{figure}[!t]
\centering
\subfloat[][]{\includegraphics[width=1.3in]{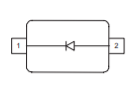}
\label{fig_first_case}}
\subfloat[]{\includegraphics[width=1.5in]{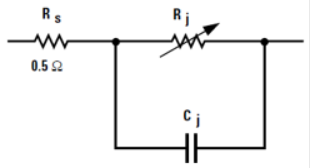}
\label{fig_second_case}}

\caption{PIN diode (a) Schematic (b) Equivalent circuit}
\label{fig:foobar}
\end{figure}

\subsection{Operation Mechanism}
In this part, polarization of proposed antenna according to excited ports of the branch line coupler is discussed. Polarization control is undertaken by two single PIN diodes (BAR64-02w) (Fig. 4). These PIN diodes are placed in the output ports of the Wilkinson power divider which are connected to input ports of the branch line coupler in final designed structure of the antenna Fig. 5. PIN diodes control the RF signal flow toward the branch line coupler determining exciting phase and amplitude of the antenna. The equivalent circuit of the PIN diode according to company’s datasheet is depicted in Fig. 4 [18]. The variable resistor (Rj) determines the ON or OFF states of the PIN diode controlled by the DC biasing circuit. 
As a result, biasing current provides an access to ON and OFF states of PIN diodes by changing variable resistor. Basically, according to equivalent circuit of the PIN diode, PIN diode switching circuit is realized in two configurations: (1) Series, (2) Shunt \cite{r32}. In this paper, shunt configuration of PIN diode with biasing circuit in Fig. 6 is utilized. A constant 45-ohm resistor is fixed at the input port of both switching circuits. Switching circuit can provide just over -3dB insertion loss in OFF state and -14 dB isolation in ON state which is appropriate for adjusting the RF signal flow which is applied to square patch antenna and achieving different operation modes. These two switching circuits are located at the input ports of branch line coupler. Consequently, according to the configuration of the switches, if two PIN diodes are OFF, both branch line coupler’s inputs are excited by the applied RF signal which lead to same phases and amplitude in inputs of the square patch antenna. So, linear polarization will be radiated at designed frequency. Besides, if diode 1 is OFF and 2 is ON, only one input of the branch line coupler is excited which makes phase difference between two excited modes of the square patch antenna. So, in this operation mode Right Hand Circular Polarization (RHCP) is radiated at designed frequency. Also this structure can radiate Left Hand Circular Polarization (LHCP). If diode 1 is ON and diode 2 is OFF, unlike the previous state, phase difference and identical amplitude are prepared reversely on input ports of the branch line coupler which lead to LHCP in this operation mode. 
As a result, the antenna can radiate three different polarizations with one feed port on the same structure. Operation modes of the antenna with states of PIN diodes are shown in Table \Romannum{4}.

\begin{figure}[!t]
\centering
\includegraphics[width=2.5in]{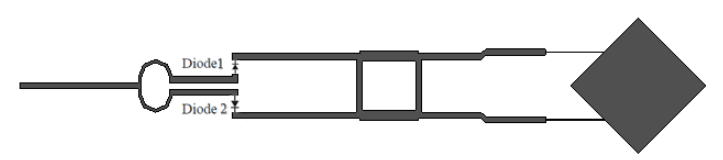}
\caption{Final structure of proposed antenna}
\label{fig_sim}
\end{figure}

\begin{figure}[!t]
\centering
\includegraphics[width=2.5in]{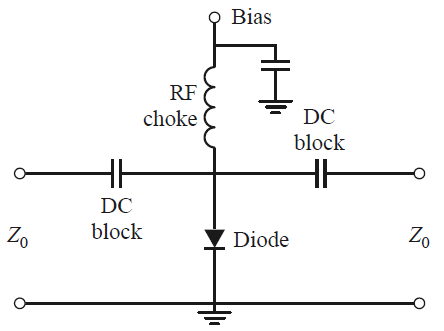}
\caption{Schematic of switching circuit \cite{r32}}
\label{fig_sim}
\end{figure}

\begin{table}[!t]
\renewcommand{\arraystretch}{1.3}
\caption{Antenna Polarization}
\label{table_example}
\centering
\begin{tabular}{|c|c|c|c|c|}
\hline
Polarization & LP (Ant.1) & RHCP (Ant.2) & LHCP (Ant.3) \\
\hline
Diode 1 & OFF & OFF & ON \\
\hline
Diode 2 & OFF & ON & OFF \\
\hline
\end{tabular}
\end{table}

\subsection{DC Biasing}
 A shunt DC biasing circuit shown in Fig. 6 is utilized in the switching circuit of this reconfigurable patch antenna. Two 47nF capacitors in the 0402 standard size package are used in the RF input and output of the switching circuit. These capacitors prevent leaking DC voltage to the RF source. An external voltage supply makes a 5mA current to change the state of the PIN diode. This voltage is applied through a 22nH inductance in the 0402 standard size package blocking RF leakage into the DC voltage supply. 
The voltage supply with a 47nF grounded capacitor in the 0402 standard package is connected to each switching circuit through two vias placed underneath the antenna structure and the applied voltage can be controlled by two mechanical switches which are placed separately in an external box for better access.

\begin{figure}[!t]
\centering

\includegraphics[width=2.5in]{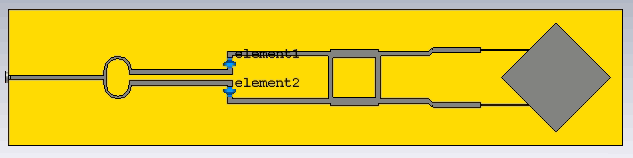}
\caption{Schematic of antenna in CST simulation}
\label{fig_sim}
\end{figure}

\section{Simulation Result}

\begin{figure}[!t]
\centering
\includegraphics[width=3in]{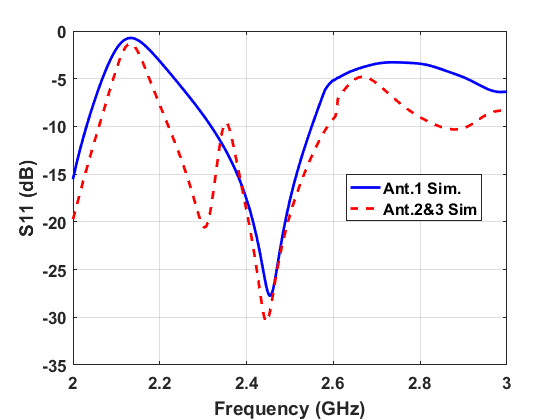}
\caption{Simulated return loss for Ant.1}
\label{fig_sim}
\end{figure}

\begin{figure}[t]
\centering
\includegraphics[width=3in]{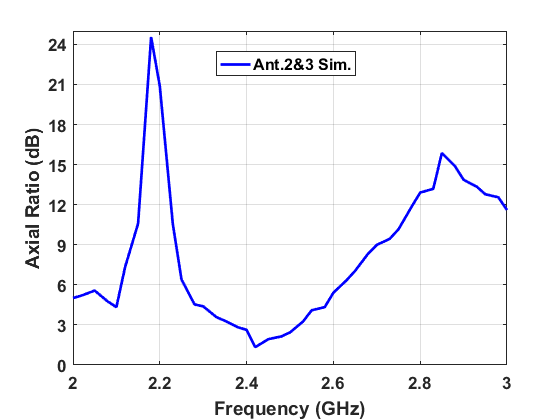}
\caption{Simulated axial ratio for Ant.2 and Ant.3}
\label{fig_sim}
\end{figure}

\begin{figure}[!t]
\centering
\subfloat[][Linear polarization for (Ant.1)]{\includegraphics[width=2.5in]{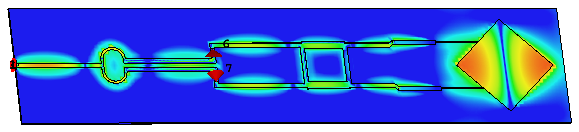}
\label{fig_first_case}}

\subfloat[Right hand circular polarization (RHCP) for (Ant.2)]{\includegraphics[width=2.5in]{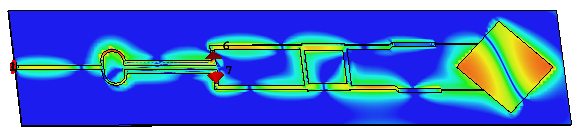}
\label{fig_second_case}}

\subfloat[Left hand circular polarization (LHCP) for (Ant.3)]{\includegraphics[width=2.5in]{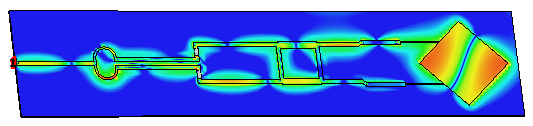}
\label{fig_second_case}}

\caption{Simulated electrical field distribution for all operation modes}
\label{fig:foobar}
\end{figure}

The primary implementation of the discussed structure in the previous section is performed and simulated in the layout part of the Advanced Design System (ADS) software then it is converted to the CST software (Fig. 7) for validating and checking radiation characteristics. DC analysis and simulation of switching circuits in ON and OFF states which was mentioned in B. Operating mechanism is done by the ADC software then the touchstone file is exported to be utilized in further simulation. In the CST simulation, the aforementioned touchstone files of switching circuits for all operation modes of antenna are placed in the schematic part of the CST (Fig.7) to attain radiation characteristics in CST results. The simulated return loss of Ant.1 is shown in Fig 8. The minimum simulated return loss in the linear polarization state (Ant.1) is -27.7 dB at 2.45 GHz and the impedance bandwidth is 9.5\%. The return loss of Ant.2 and Ant.3 are drawn in Fig. 8. these reflection coefficients are equal since the structure of the antenna is symmetrical. The minimum simulated return loss in the right-hand circular polarization RHCP (Ant.2) and the left-hand circular polarization state (Ant.3) is -30.4 dB at 2.44 GHz and the impedance bandwidth is 9\%. As a result, the operating frequency is identical for all antennas on the same structure. Moreover, simulated impedance bandwidth in all 3 antennas is 200 MHz (9\%) which is acceptable for wireless applications. Radiating circular polarization is necessarily having bellow 3 dB axial ratio (AR) in both right hand (Ant.2) or left hand (Ant.3) circular polarization. The simulated axial ratio (AR) for Ant.2 and Ant.3 is drawn in Fig. 9. Due to the symmetrical structure of this antenna, both antennas have the same axial ratio. The minimum simulated axial ratio is 1 dB at 2.43GHz and the axial ratio bandwidth (ARBW) is 4.5\%. What determines the polarization of the antenna which can be controlled by switching circuits is adjusting the form of current distribution. It can make 0 and 90 degrees phase difference in E-field distribution which is applied to square patch and causes linear polarization and circular polarization respectively. The E-field distributions related to each antenna are depicted in Fig. 10 (a)-(c). Optically, the same phase excitation is seen in Fig. 10 (a) which is related to linear polarization Ant.1. In addition, 90 degrees phase difference in right-hand circular polarization (RHCP) and left-hand circular polarization (LHCP) is extracted in Fig. 12 (b) and 12 (c), respectively.

\section{Experimental Result}

\begin{figure}[!t]
\centering
\includegraphics[width=2in]{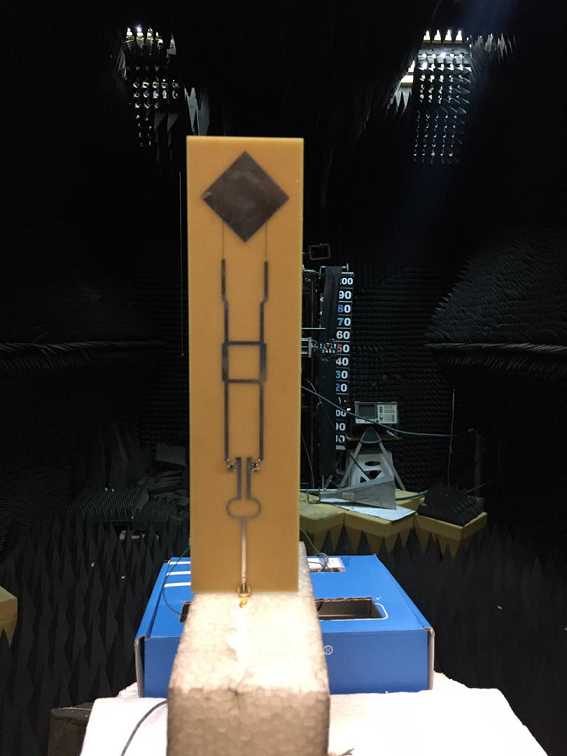}
\caption{Antenna in testing conditions}
\label{fig_sim}
\end{figure}

\begin{figure}[!t]
\centering
\includegraphics[width=3.2in]{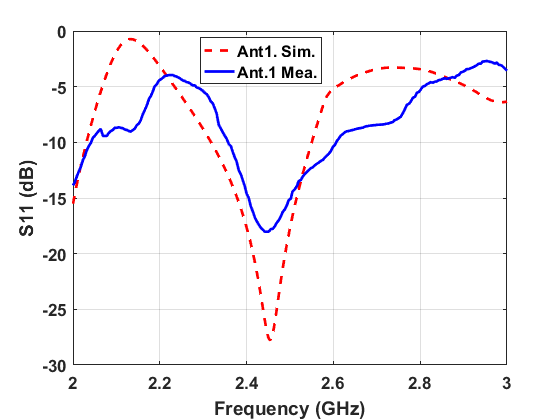}
\caption{Simulated and measured return loss for Ant.1}
\label{fig_sim}
\end{figure}

\begin{figure}[!t]
\centering
\includegraphics[width=3.2in]{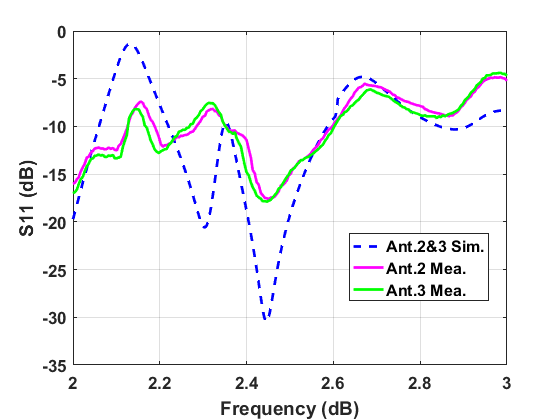}
\caption{Simulated and measured return loss for Ant.2 and 3}
\label{fig_sim}
\end{figure}

\begin{figure}[!t]
\centering
\includegraphics[width=3.2in]{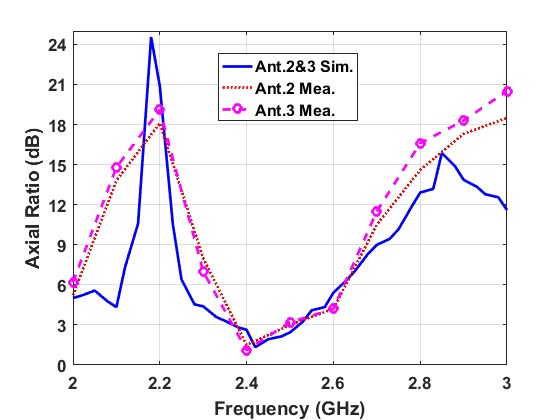}

\caption{Simulated and measured axial ratio for Ant.2 and Ant.3}
\label{fig_sim}
\end{figure}
In this section, experimental and simulation results will be discussed. Fig. 11 exhibits the designed antenna prototype in testing conditions fabricated by the printed circuit board (PCB) technology and its external DC controlling switches.

\subsection{Matching}
The measured and simulated reflection coefficient for Ant.1 is drawn in Fig. 12. Linear polarization in both results has the same -10dB bandwidth and operating frequency. The -10dB bandwidth for simulated is 8\% and the minimum return loss in this antenna is at 2.45 GH. Moreover, measured and simulated reflection coefficient for RHCP(Ant.2) and LHCP(Ant.3) state of this antenna are depicted in Fig. 13. Return loss in these circular polarization operation modes of this antenna has the same -10-dB bandwidth and operating frequency of 8\% and 2.45GHz respectively. Consequently, comparing measured return loss of all possible states of this antenna implies that this antenna has satisfactory impedance matching in all possible states at designed frequency.
\subsection{Radiation Characteristics}
The measured and the simulated axial ratio of Ant.2 (RHCP) and Ant.3 (LHCP) are illustrated in Fig. 14. All measured results are done in the main lobe of the antenna pattern (theta=30, Phi=0) and 200 MHz steps are used for obtaining measured results. The measured axial ratio bandwidth (ARBW) of Ant.2 and Ant.3 are (2.39-2.48 GHz) 3.63\% and (2.36-2.47 GHz) 4.4\%, respectively. Thus, this antenna can radiate right-hand and left-hand circular polarization on the same structure at designed frequency. Likewise, the small discrepancies between the simulated and measured are rooted in the small tolerance of the fabrication process. As a result of the axial ratio and reflection coefficient, this antenna can perform at 3 different polarizations in a specific frequency according to the overlap of axial ratio bandwidth and -10 dB bandwidth in all possible states of the antenna.

Normalized radiation patterns of this antenna in all operation modes are illustrated in Fig. 15. Measured radiation patterns in E-plane and H-plane for LHCP and RHCP modes of this antenna are approximately equal due to the symmetrical antenna structure. All directional patterns of operation modes are achieved with 1.2 dBi, 1 dBi, and 0.98 dBi gain for Ant.1, Ant.2, and Ant.3, respectively.
Consequently, this antenna can perform at three different polarization at specific frequency according to the overlap of axial ratio bandwidth and -10 dB bandwidth in all possible states of the antenna.

\section{Conclusion}

\begin{figure}[!t]
\centering
\subfloat[][]{\includegraphics[width=1.8in]{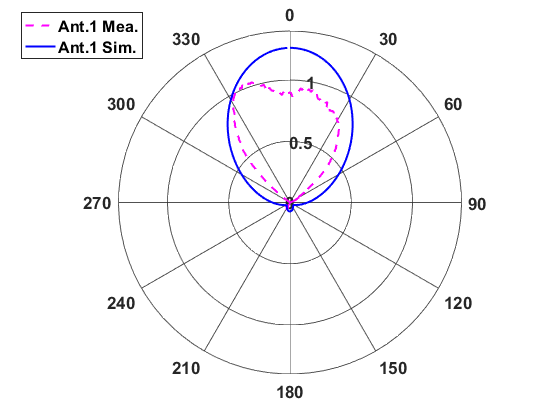}
\label{fig_first_case}}
\subfloat[]{\includegraphics[width=1.8in]{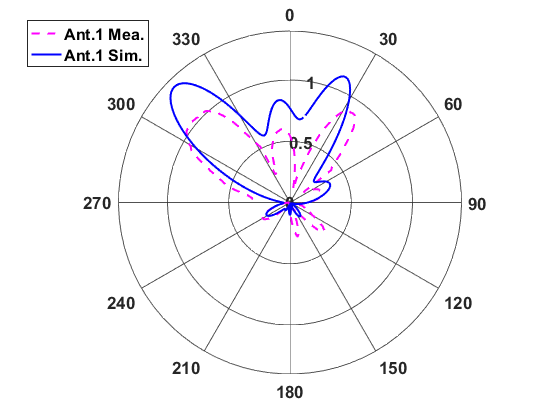}
\label{fig_second_case}}

\subfloat[]{\includegraphics[width=1.8in]{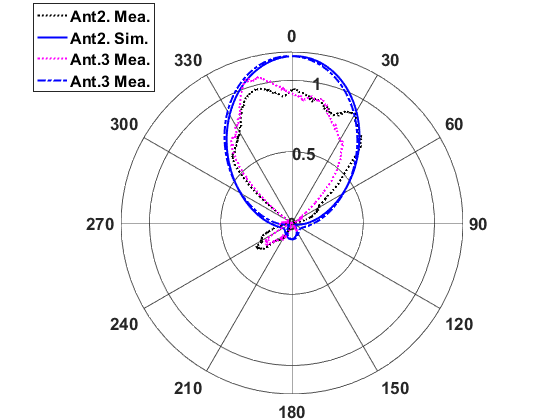}
\label{fig_second_case}}
\subfloat[]{\includegraphics[width=1.8in]{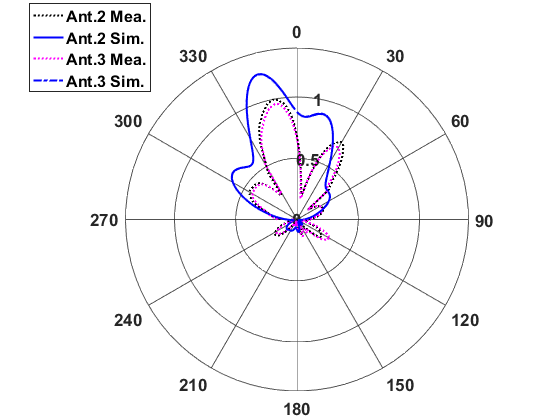}
\label{fig_second_case}}

\caption{Simulated and measured radiation pattern at 2.45 GHz of (a) LP in H-Plane (b) LP in E-Plane (c) RHCP and LHCP in H-Plane (d) RHCP and LHCP in E-Plane }
\label{fig:foobar}
\end{figure}

A triple-polarization square patch antenna has been presented in this paper. Polarization reconfigurability with the same feeding port is achieved by a combination of two power dividers and two PIN diodes as switching circuits. These switching circuits control the RF signal flow applied to the power dividers and the square patch antenna. Additionally, polarization switching can be handled by changing the DC voltage of the shunt biasing circuits of PIN diodes. The impedance bandwidth coincides with the axial ratio’s bandwidth according to measured results. So, this design considers all needs for wireless applications at 2.45 GHz.

\ifCLASSOPTIONcaptionsoff
  \newpage
\fi

\bibliographystyle{IEEEtran}
\bibliography{ref}

\end{document}